\documentclass[12pt,aps,prd,showpacs,amsmath,amssymb,floatfix,nofootinbib,superscriptaddress]{revtex4-1}

\usepackage{graphicx}  % needed for figures
\usepackage{multirow}
\usepackage[dvipsnames]{xcolor}
\usepackage[colorlinks=true,allcolors=BlueViolet]{hyperref}
\usepackage{latexsym}
\usepackage{amsfonts}
\usepackage{amsmath,amssymb}
\usepackage{slashed}
\usepackage{color}

\usepackage{supertabular} 
\usepackage{epsfig}
\usepackage{lipsum}

\newcommand{\mpo}{m_{p_1}}

\newcommand{\mpw}{m_{p_2}}

\newcommand{\mko}{m_{k_1}}

\newcommand{\mkw}{m_{k_2}}

\newcommand{\Tr}{\text{Tr}}
\newcommand{\fxg}{\gamma g^2}
\newcommand{\fxgg}{\gamma^2 g^2}
\newcommand{\fxgp}{\gamma^{\prime} g^2}
\newcommand{\fxggp}{\gamma^{\prime2} g^2}
\newcommand{\fxgpg}{\gamma\gamma^{\prime} g^2}

\newcommand{\uestc}{\affiliation{School of Physics, University of Electronic Science and Technology of China, Chengdu 611731, China}}

\newcommand{\ific}{\affiliation{Instituto de F\'isica Corpuscular (centro mixto CSIC-UV), \\
Institutos de Investigaci\'on de Paterna, Apartado 22085, 46071, Valencia, Spain}}

\newcommand{\uv}
{\affiliation{Departamento de F\'{\i}sica Te\'orica and IFIC, Centro Mixto Universidad de Valencia-CSIC Institutos de Investigaci\'on de Paterna, Aptdo.22085, 46071 Valencia, Spain}
}

\begin{document}

\title{Pole analysis for the \boldmath $D^{*}\bar K$-$D\bar{K^*}$ coupled-channel system}

\author{Pan-Pan~Shi} \email{Panpan.Shi@ific.uv.es}
\uv\ific

\author{F. Gil-Dom\'inguez}
\email{fgildominguez@uestc.edu.cn}
\uv
\uestc
\author{R. Molina}
\email{Raquel.Molina@ific.uv.es}
\uv

\author{Meng-Lin~Du}
\email{du.ml@uestc.edu.cn}
\uestc

% \date{\today}

\begin{abstract}

By solving the Lippmann-Schwinger equation, possible hadronic molecules in the $D^*\bar K$-$D\bar K^*$ coupled-channel system are investigated with the one-meson exchange potentials, where both vector and pseudoscalar mesons are considered as exchange particles. We find an S-wave virtual state with mass $M=2487$~MeV, and a resonance with $M=2759$ and width $\Gamma=18$~MeV. In the $D^* \bar K$ invariant mass distribution, the virtual state appears as a cusp at the $D^*\bar K$ threshold, while the resonance potentially manifests as a dip. In particular, we take into account the $D\bar K \pi$ three-body dynamics due to the on-shell pion exchange and the finite decay width for $\bar K^*$.
\end{abstract}

\maketitle

\section{Introduction}

A large number of hadronic structures have been observed in the last two decades. Many of these structures exhibit properties that are incompatible with the predictions of the traditional quark model, which describes the ordinary hadrons as quark-antiquark pairs (mesons) or three quarks (baryons). Since QCD does not prevent the formation of more complicated structures, e.g. tetraquark and pentaquark states, those recently observed states are candidates for exotics.  See Refs.~\cite{
Chen:2016qju,Hosaka:2016pey,Esposito:2016noz,Lebed:2016hpi,Ali:2017jda,Olsen:2017bmm,Guo:2017jvc,Albuquerque:2018jkn,Liu:2019zoy,Guo:2019twa,Brambilla:2019esw,Chen:2022asf,Liu:2024uxn} for recent reviews. In particular, hadrons with exotic quantum numbers should be excellent candidates for exotic states. For instance, the $T_{cc}^+$ is observed in the $D^0D^0\pi^+$ invariant mass distribution~\cite{LHCb:2021auc,LHCb:2021vvq} and its quark configuration indicates it is an ideal double-charm exotic candidate.

Focusing on the open-charm and strange sector, the $D^*_{s0}(2317)$~\cite{BaBar:2003oey} and $D_{s1}(2460)$~\cite{CLEO:2003ggt} were first observed by the BaBar and CLEO Collaborations. Their masses are significantly lower than the scalar and axial-vector $c\bar s$ mesons predicted by the quark model, see, e.g., Ref.~\cite{Godfrey:1985xj}. Because of the closeness of their masses to the $DK$ and $D^*K$ thresholds, these are more likely exotic candidates. See recent reviews, Refs.~\cite{Olsen:2017bmm,Guo:2017jvc}. Later, the LHCb Collaboration reported several exotic candidates with unconventional quantum numbers in $B$ meson decay processes. In 2020, two structures, the $X_{0}(2900)$ and $X_{1}(2900)$, were observed in the $D^-K^+$ invariant mass distribution, with spin-0 and spin-1, respectively~\cite{LHCb:2020bls,LHCb:2020pxc}. Those states involve exotic quark configuration, comprising both {$\bar c$} and {$\bar s$} quarks. Subsequently, in 2023, the LHCb Collaboration identified two new structures $T_{c\bar s}^0 (2900)$ and $T_{c\bar s}^{++} (2900)$ in the $D_s^+\pi^-$ and $D_s^+\pi^+$ invariant mass distributions, with quantum numbers $J^{P}=0^+$~\cite{LHCb:2022sfr,LHCb:2022lzp}. Their decay modes suggest that these are tetraquark states with the quark content $c \bar s d \bar u$, $c \bar s u \bar d$, respectively. 

Such an abundance of structures observed in the open-charm sector has spurred further exploration into possible exotic states within the charm-strange system. To understand the nature of those states, various phenomenological models are utilized to explore these structures. There are several investigations for the $X_0(2900)$. Based on the one-boson exchange model, an $S$-wave $D^* \bar K^*$ molecular state, which is associated with $X_0(2900)$ via the charge conjugation, was predicted in Ref.~\cite{Molina:2010tx}. After the LHCb Collaboration reported the $X_0(2900)$, this state is explored as the $\bar D^* K^*$ molecule in terms of the one-meson exchange potentials in Refs.~\cite{Molina:2020hde,He:2020btl,Kong:2021ohg,Ke:2022ocs,Liu:2020nil}, contact potentials in Ref.~\cite{Wang:2023hpp}, and the QCD sum rules in Refs.~\cite{Agaev:2022duz,Agaev:2022eyk,Chen:2020aos,Agaev:2020nrc}. Besides, $X_0(2900)$ is also interpreted as  tetraquark-like structures within the constituent quark model~\cite{Ortega:2023azl}, and the compact tetraquark state~\cite{Wang:2020xyc,Wei:2022wtr,Liu:2022hbk,Zhang:2020oze,Lu:2020qmp,Agaev:2022eeh,He:2020jna}.
Spin partners of the $X_0(2900)$ are also predicted~\cite{Molina:2010tx,Molina:2020hde}, and several reactions are proposed to search for them~\cite{Dai:2022qwh, Dai:2022htx}. The $X_1(2900)$ could be considered as  an $S$-wave $\bar D_1K$ virtual state~\cite{He:2020btl}, a $P$-wave hadronic molecule~\cite{Wang:2021lwy} and the compact vector tetraquark state~\cite{Chen:2020aos,Agaev:2021knl,Wei:2022wtr,He:2020jna}. The kinematic effect is also explored to explain the structures of $X_0(2900)$ and $X_1(2900)$, where the triangle singularities enhance the rescattering of $D^{*-}K^{*+}\to D^-K^+$ and $\bar D^0_1 K^0\to D^-K^+$ and lead to the production of two resonance-like structures~\cite{Liu:2020orv}. Regarding the $T_{c\bar{s}}(2900)$, the cusp explanation from the $D^*K^*$-$D_s^*\rho$ coupled channel reproduces well the invariant mass distribution in the decay mode $B \to \bar{D} D_s\pi$~\cite{Molina:2022jcd}. In~\cite{Duan:2023lcj} the structure is attributed to a bound state or a virtual state which manifests as the cusp in \cite{Molina:2022jcd}. Besides, $T_{c\bar s}(2900)$ is also explored as compact tetraquark states~\cite{Nunavath:2022tgv,Liu:2022hbk,Yang:2023wgu,Ortega:2023azl,Li:2024ctd}. See also~\cite{Yang:2023evp,Ortega:2023azl,Duan:2023qsg,Wang:2023hpp} and the recent review on new hadron states~\cite{Chen:2022asf}. These new discoveries in the open charm and strange sector point out the existence of tetraquarks in nature. 

After these observations, it is natural to think that similar states close to the  $D^*\bar{K}$, $D\bar{K}^*$ thresholds should exist. Still, only a few hadronic molecules of this kind have been predicted. Using an extension of the SU(3) chiral Lagrangian to SU(4), but explicitly breaking the symmetry in the masses and couplings, in Ref.~\cite{Gamermann:2007fi}, the $D^*\bar K$-$D\bar K^*$ coupled-channel system is studied. As a result, in the $D^*\bar K$-$D\bar K^*$ coupled-channel system, a narrow pole (located at $2756.08-2.15i$ MeV) is predicted, which is close to the $D\bar K^*$ threshold, and a broad cusp is found around the $D^*\bar K$ threshold. For the narrow pole, its width is significantly impacted by the $\bar K^*$ decay width once taking into account the contribution of the finite width of $\bar K^*$ meson in the Green's function.

In the local-hidden-gauge approach, the isoscalar $D^*\bar K^*$ molecule was predicted with spins $J=0,~1,$ and $2$~\cite{Molina:2010tx,Molina:2020hde}, before its discovery, being consistent with the recent observation of the $T_{cs}(2900)$, while the $T_{c\bar{s}}(2900)$ can be reproduced by a cusp~\cite{Molina:2010tx,Molina:2022jcd} which is attributed to a virtual state \cite{Duan:2023lcj}.
These recent discoveries reinforce the predictions of~\cite{Gamermann:2007fi} regarding the predictions of dynamically generated resonances from the $D^*\bar K$-$D\bar K^*$ coupled-channel system. Therefore, it is necessary to revisit this topic. However, note that in these previous works, only the vector meson exchange is considered. In particular, since the $D \bar K^*$ channel can be connected to the $D^*\bar K$ channel through the exchange of the $\pi$, $\eta$, and $\eta^{\prime}$ mesons, there can be some effect due to the light pseudoscalar mesons in the narrow pole predicted in~\cite{Gamermann:2007fi},
where only the $D_s^*$-exchange can connect these channels. Besides, {$D^*$ and $\bar K^*$ are unstable and their decay widths}, not considered in~\cite{Gamermann:2007fi}, cannot be overlooked. Their decay processes, especially associated with the $D\bar K \pi$ three-body effect, could play a significant role in the dynamics of this system. In addition, the $D\bar K \pi$ three-body threshold is very close to the $D^*\bar K$ threshold. Hence, the $K^*$ and $D^*$ decay widths could have a significant influence on the overall behavior of poles.

To address these issues comprehensively, we employ the complete formula for the meson exchange potentials and solve the integral Lippmann-Schwinger (LS) equation with the use of the effective Lagrangian approach. Thus, our approach incorporates the contributions from the vector and pseudoscalar meson exchange for the $D^*\bar K$-$D\bar K^*$ coupled-channel system, as well as, the decay widths of $D^*$ and $\bar K^*$, to $D\pi$ and $\bar{K}\pi$, that leads to three-body intermediate $D\bar{K}\pi$ states. However, since the contribution of the $D^*$ decay width is more than two orders of magnitude smaller than the $\bar K^*$  width, the $D^*$ decay width can be safely neglected in our calculation. Based on that framework, we aim to predict the possible hadronic molecules and examine their coupling with relevant channels. {Note that we neglect the contribution of the $D^*\bar K^*$ channel due to the fact that its coupling to the $D^*\bar K$--$D\bar K^*$ system is very small since it goes through a vector-vector-pseudoscalar anomalous vertex, which is $\mathcal{O}(p^2)$~\cite{Meissner:1987ge}, as well as there is a large threshold gap between this channel and the $D^* \bar K$ threshold of around $400$~MeV.    }

This paper is organized as follows. In Sect. \ref{Sec:formula}, we present the effective potential for the $D^*\bar K$-$D\bar K^*$ coupled-channel
system and the three-body dynamics. The numerical results, including the prediction of possible molecules and the {line shapes}, are discussed in Sect.~\ref{sec:result}. A brief summary is given in Sect.~\ref{sec:sum}. We discuss the details for the partial wave projection of the effective potentials in  Appendix~\ref{Sec:partial_wave}. {In  Appendix~\ref{Sec:Dstar_decay}, we discuss the decay formula for $D^*\to D\pi$ within the frameworks of the SU(4) symmetry and Heavy Quark Symmetry.}

\section{Formalism}\label{Sec:formula}

The interaction for the $D^*\bar K$-$D\bar K^*$ can be described by the local hidden gauge formalism~\cite{Bando:1984ej,Bando:1987br,Meissner:1987ge,Nagahiro:2008cv,Molina:2009ct,Molina:2010tx}. Within this formalism, the interactions among vector mesons and between vector meson and pseudoscalar mesons, can be described by the hidden-gauge-symmetry  Lagrangian~\cite{Nagahiro:2008cv,Molina:2009ct,Molina:2010tx}
\begin{align}
\mathcal{L}_{I I I}^{(3 V)}&=i g\Tr\left[\left(\partial_\mu V_\nu-\partial_\nu V_\mu\right) V^\mu V^\nu\right],\nonumber\\
\mathcal{L}^{\phi\phi V}&=i g \operatorname{Tr}\left(\left[\partial_\mu \phi, \phi\right] V^\mu\right).
\label{Eq:Lagrangian_SU4}
\end{align}
The coupling constant $g$ is defined as
\begin{align}\label{eq:coup}
    g=\frac{m_{\rho}}{2F_{\pi}}\ ,
\end{align}
where $F_{\pi}=93$ MeV is the pion decay constant, and $m_{\rho}$ is the mass of $\rho$ meson\footnote{Note that we use the isospin average mass for the mesons discussed in this work.}. The matrices for the vector mesons $V_{\mu}$ and pseudoscalar mesons $\phi$ are, respectively,
\begin{align}\label{eq:v}
V_\mu=\left(\begin{array}{cccc}
\frac{\rho^0}{\sqrt{2}}+\frac{\omega}{\sqrt{2}} & \rho^{+} & K^{*+} & \bar{D}^{* 0} \\
\rho^{-} & -\frac{\rho^0}{\sqrt{2}}+\frac{\omega}{\sqrt{2}} & K^{* 0} & D^{*-} \\
K^{*-} & \bar{K}^{* 0} & \phi & D_s^{*-} \\
D^{* 0} & D^{*+} & D_s^{*+} & J / \psi
\end{array}\right)_\mu, 
\end{align}
and
\begin{align}\label{eq:p}
\phi=\left(\begin{array}{cccc}
\frac{\pi^0}{\sqrt{2}}+\frac{\eta}{\sqrt{3}}+\frac{\eta^{\prime}}{\sqrt{6}} & \pi^{+} & K^{+} & \bar{D}^0 \\
\pi^{-} & -\frac{\pi^0}{\sqrt{2}}+\frac{\eta}{\sqrt{3}}+\frac{\eta^{\prime}}{\sqrt{6}} & K^0 & D^{-} \\
K^{-} & \bar{K}^0 & -\frac{\eta}{\sqrt{3}}+\sqrt{\frac{2}{3}} \eta^{\prime} & D_s^{-} \\
D^0 & D^{+} & D_s^{+} & \eta_c
\end{array}\right).
\end{align}
The interaction can be calculated from the Feynman diagrams depicted in Fig.~\ref{Fig:diagram}, where we consider explicitly light and heavy meson exchanges for both pseudoscalar and vector mesons. {In contrast to the $NN$ and other baryon-baryon systems~\cite{Oset:2000gn,Haidenbauer:2005zh,Wu:2024trh}, for the $D\bar D^*$, $D^*\bar D^{*}$, $B\bar B^*$, and $B^*\bar B^*$ systems, the interaction mediated by the $\sigma$-meson exchange is very weak~\cite{Aceti:2014kja,Aceti:2014uea,Dias:2014pva}. Therefore, the contribution from the $\sigma$-meson exchange is neglected in our calculation. }

\begin{figure}[tbh]
    \includegraphics[height=0.2\columnwidth]{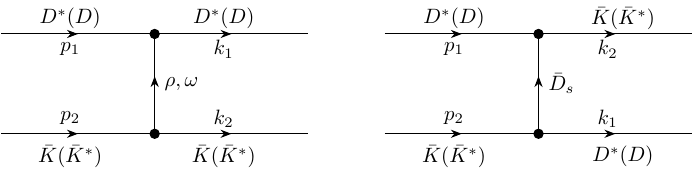} \\[10pt]
    \includegraphics[height=0.2\columnwidth]{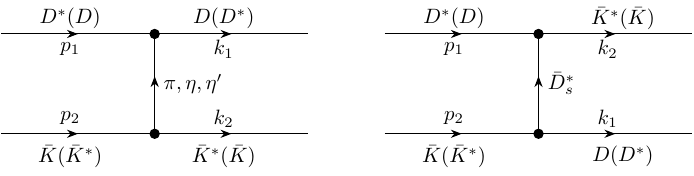}
    \caption{Feynman diagrams for the interaction between the charmed and strange mesons.}
    \label{Fig:diagram}
\end{figure}

Note that, even though SU(4) symmetry is assumed in the vertices through Eqs.~(\ref{eq:v}) and (\ref{eq:p}), in practice, this is broken to an SU(3) symmetry in the dominant terms, where there is a light meson exchange, since, in these terms, the charm quark is acting as a spectator, consistently with heavy quark spin symmetry~\cite{Xiao:2013yca}. 
The tree-level amplitudes in the $D^{*}\bar{K}$-$D\bar K^*$ system are derived from the effective Lagrangian of Eq.~\eqref{Eq:Lagrangian_SU4}. The isoscalar potentials for the $D^*\bar{K}$-$D\bar K^*$ system are
\begin{align}
{V}_{11}=& -\frac{1}{2}\fxg\left[3D_{su}(m_{\rho})-D_{su}(m_{\omega})\right]\varepsilon(p_1)\cdot\varepsilon^*(k_1)-4\fxggp D_u(m_{D_s}){k_2\cdot\varepsilon(p_1)p_2\cdot\varepsilon^*(k_1)},\nonumber\\
V_{12} =& -\fxgpg D_{st}(m_{D_s^*})\varepsilon(p_1)\cdot\varepsilon^*(k_2)+\frac{2\fxgp}{3}k_1\cdot\varepsilon(p_1)p_2\cdot\varepsilon^*(k_2)\left(4D_t(m_{\eta})-D_t(m_{\eta'})-9D_{\pi}\right),\nonumber\\
V_{21} =& -\fxgpg D_{st}(m_{D_s^*})\varepsilon(p_2)\cdot\varepsilon^*(k_1)+\frac{2\fxgp}{3}k_2\cdot\varepsilon(p_2)p_1\cdot\varepsilon^*(k_1)\left(4D_t(m_{\eta})-D_t(m_{\eta'})-9D_{\pi}\right),\nonumber\\
{V}_{22}=&  -\frac{1}{2}\fxg\left[3D_{su}(m_{\rho})-D_{su}(m_{\omega})\right]\varepsilon(p_2)\cdot\varepsilon^*(k_2)-4\fxgg D_u(m_{D_s})k_1\cdot\varepsilon(p_2)p_1\cdot\varepsilon^*(k_2),
\label{Eq:potential_OME}
\end{align}
where the subscripts of $V_{ij}$ denote the $i$-th and $j$-th channels, $p_{\alpha}$ and $k_{\alpha}$ are the momenta of initial and final states ($\alpha=1$ for charmed mesons and $\alpha=2$ for light mesons), and $\varepsilon(q)$ is the polarization vector of the vector mesons with momentum $q$. Moreover, the formal SU(4) symmetry used in the vertices through Eqs.~(\ref{eq:v}) and (\ref{eq:p}), is not only broken in the masses, which are taken from the PDG~\cite{ParticleDataGroup:2022pth}, but we also allow for breaking in the couplings by inserting the breaking parameters $\gamma$ and $\gamma^{\prime}$. These are introduced here to explore the SU(4) breaking effect of the global coupling $g$ in Eq.~(\ref{eq:coup}) between the heavy and light mesons coming from the effective Lagrangians $\mathcal{L}_{I I I}^{(3 V)}$ and $\mathcal{L}^{(\phi\phi V)}$. {For the interaction of only light mesons, both the breaking parameters $\gamma$ and $\gamma^{\prime}$ is set to one. {Note that, in this way, with the Lagrangian given in Eq.~(\ref{Eq:Lagrangian_SU4}) and the value of $g$ from  Eq.~(\ref{eq:coup}) the $K^*$ experimental decay width is well reproduced~\cite{Tolos:2010fq}.}} We note that the exponential form factor is introduced to account for the pion exchange~\cite{Navarra:2001ju,Molina:2020hde,Molina:2010tx,Molina:2020hde,Molina:2022jcd}. In this work, however, we simply use a constant $\gamma^{\prime}$ to parametrize this difference. Thus, we will study the dependence of the results in the next section with these parameters. In the above equation, the terms without polarization vectors are defined as
\begin{align}
D_{su}(m)&=\frac{s-u}{t-m^2};\quad
D_{t}(m)=\frac{1}{t-m^2};\nonumber\\
D_{st}(m)&=\frac{s-t}{u-m^2};\quad
D_{u}(m)=\frac{1}{u-m^2},
\end{align}
with the Mandelstam variables $s=(p_1+p_2)^2,~t=(p_1-k_1)^2$, and $u=(p_1-k_2)^2$.
In our calculation, the pion propagator $D_{\pi}$ is decomposed into two parts in terms of the time-ordered perturbation theory (TOPT) \cite{Lensky:2005hb,Baru:2004kw,Baru:2009tx,Du:2021zzh}
\begin{align}
D_{\pi}
=\frac{1}{2E_{\pi}}\left[\frac{1}{\sqrt{s}-E_{p_2}-E_{k_1}-E_{\pi}+i\varepsilon}+\frac{1}{\sqrt{s}-E_{p_1}-E_{k_2}-E_{\pi}+i\varepsilon}\right],
\end{align}
where $E_{\pi}=\sqrt{m_{\pi}^2+(\vec{k}_1-\vec{p}_1)^2}$. 

To account for the partial wave contributions, we decompose the $D^*\bar K$-$D\bar K^*$ system into four channels
\begin{align}
\left\{D^*\bar K(S),~D^*\bar K(D),~D\bar K^*(S),~D\bar K^*(D) \right\},
\label{Eq:channel}
\end{align}
where $S$ and $D$ represent the $S$- and $D$-wave components, respectively. We then rewrite the effective potential matrix as
\begin{align}
V=\left(\begin{array}{cccc}
V_{11}^{SS} & V_{11}^{SD} & V_{12}^{SS} & V_{12}^{SD}  \\
V_{11}^{DS} & V_{11}^{DD}& V_{12}^{DS} & V_{12}^{DD}  \\
V_{21}^{SS} & V_{21}^{SD} & V_{22}^{SS} & V_{22}^{SD}  \\
V_{21}^{DS} & V_{21}^{DD} & V_{22}^{DS} & V_{22}^{DD}  \\
\end{array}\right),
\label{Eq:potential}
\end{align}
where the elements of the potential matrix can be derived from the partial-wave projection of the effective potentials in Eq.~\eqref{Eq:potential_OME}. Further details on the evaluation of these coefficients are discussed in Appendix~\ref{Sec:partial_wave}.

The relativistic Green's function in our calculation is given by
\begin{align}
I(E,q)=\frac{\omega_1(q)+\omega_2(q)}{2\omega_1(q)\omega_2(q)}\frac{1}{E^2-\left[\omega_1(q)+\omega_2(q)\right]^2+im_{\alpha}\Gamma_{\alpha}(M)},
\label{Eq:green}
\end{align}
where $w_{\alpha}(q)=\sqrt{m_{\alpha}^2+q^2}$ and $\Gamma_{\alpha}(M)$ denotes the decay width of $\alpha$-th hadron. Here we employ a momentum-dependent decay width, which is equivalent to the self-energy contribution of the unstable hadrons.  The decay width of $D^*$, where total {widths} for $D^{*0}$ and $D^{*+}$ are 56.2 and 83.4 keV~\cite{ParticleDataGroup:2022pth,Albaladejo:2021vln}, is much smaller than the $\bar K^{*}$ width ($51.4$ and $47.3$ MeV for charged and neutral, $\bar K^*$, respectively~\cite{ParticleDataGroup:2022pth}). Therefore, we neglect the width of $D^*$ and consider it as a stable particle. In the $D\bar K^*$ channel,  the decay {width} of $\bar K^*$ is
\begin{align}
\Gamma_{\bar K^*}(M_{\bar K^*})=\Gamma[{\bar K^*\to \bar K\pi}]\frac{m_{\bar K^*}^2}{M_{\bar K^*}^2}\left(\frac{p(M_{\bar K^*},m_{\bar K}, m_{\pi})}{p(m_{\bar K^*}, m_{\bar K}, m_{\pi}))}\right)^3\Theta(M_{\bar K^*}-m_{\bar K}-m_{\pi}),
\label{Eq:width_Kstar}
\end{align} 
where the $\bar K^*$ invariant mass is $M_{\bar K^*}^2=\left(E-\sqrt{m_{\bar D}^2+q^2}\right)^2-q^2$ with $E=\sqrt{s}$, the energy in the center-of-mass frame, and the Heaviside theta function is utilized to ensure no unphysical width appears below the anomalous threshold. The decay width and branching ratios of $\bar K^*$ are obtained from Ref.~\cite{ParticleDataGroup:2022pth}.  In our calculation, the momentum $p(M,m_1,m_2)$ is 
\begin{align}
p(M,m_1,m_2)=\frac{\sqrt{\lambda(M^2,m_{1}^2,m_{2}^2)}}{2M},
\end{align}
with $\lambda(a,b,c)=a^2+b^2+c^2-2ab-2ac-2bc$, and the $i$-th hadron mass $m_i$.

The scattering information of the $D^*\bar K$-$D\bar K^*$ coupled-channel system is described by the $T$-matrix, which can be calculated by solving the coupled-channel LS equation
\begin{align}
T_{ij}(E,p,k)=V_{ij}(E,p,k)+\sum_{k}\int_{\Lambda}\frac{d^3\vec q}{(2\pi)^3} V_{ik}(E,p,q)I_k(E,q)T_{kj}(E,q,k),
\label{Eq:T_matrix}
\end{align}
where the subscripts, $i$, $j$, and $k$, account for the channels, as defined in Eq.~\eqref{Eq:channel}. The potential matrix elements are listed in Eqs.~\eqref{Eq:potential_OME} and \eqref{Eq:potential}. Also, $I_k(E,q)$ represents the Green's function of Eq.~\eqref{Eq:green} in the $k$-th channel. Here, a hard cutoff is employed to regularize the UV divergence in the loop function.

Through analytical continuation, the complex $E$-plane can be decomposed into four Riemann Sheets (RSs) in terms of the two-body cuts, labeled by $(\pm,\pm)$, where the physical RS is $(+,+)$.\footnote{For the $i$-th channel, the symbols $+$ and $-$ denote the positive and negative imaginary parts of the on-shell momentum, respectively.} Since we consider the decay width of particles in the loop function, the two-body cut will deviate from the real axis of $E$-plane~\cite{Doring:2009yv} and the second RS can be reached by the continuation
\begin{align}
G^{\text{on},-}_k(E) = G^{\text{on},+}_k(E) + \frac{ip_k^{\text{on}}}{4\pi E}, 
\label{Eq:relation_RS}
\end{align}
where the $k$-channel on-shell Green's function in the physical RS is
\begin{align}
G^{\text{on},+}_k(E)&=\int\frac{d^3\vec q}{(2\pi)^3}I_k(E,q).
\end{align}
The on-shell momentum $p_k^{\text{on}}$ is determined by the zeros of the denominator of the Green's function~\eqref{Eq:green}.\footnote{Specifically, when we consider the decay width of the particles in the loop function, the on-shell momentum $p^{\text{on}}$ can be calculated by solving the equation $1/I(E,p^{\text{on}})=0$~\cite{Zhang:2024fxy}, where the imaginary part of $p^{\text{on}}$ is larger than zero.}
The poles of $T$-matrix can be searched within each RS.  Therefore, for the $D\bar K^*$ channel, the two-body cut is located at the complex $E$-plane and the relationship between different RSs is the standard formula in Eq.~\eqref{Eq:relation_RS}.
Besides, in our framework, the $D\bar K\pi$ three-body cut also appears in our energy region of interest, as the three-body threshold is very close to the $D^*\bar K$ threshold. To account for the unphysical RS for the three-body cut, we modify the momentum-dependent partial width
$\bar K^*$ through the analytic continuation in the complex energy plane
\begin{align}
\Gamma_i(M)\to -\Gamma_i(M)~~\text{for}~~\text{Im}(M)< 0,
\end{align}
where $\Gamma_i(M)$ represents partial width relevant to the three-body cut, and $M$ is invariant mass. For further details about three-body unitary and analytical continuation, one can see Refs.~\cite{Doring:2009yv,Mai:2017vot,Sadasivan:2021emk,Du:2021zzh,Wang:2023lia,Zhang:2024dth}.
The coupling constant between the pole and the $i$-th channel can be extracted from the residue of the $T$-matrix
\begin{align}
g_{i}g_{j}=\lim_{s\to E_r^2}(s-E_r^2)T_{ij}(E,p,k),
\label{Eq:coupling}
\end{align}
where $E_r$ is the pole position presented in the $r$-th RS.

\section{Numerical Result}\label{sec:result}

In this section, we calculate the poles of the open-charm system with quark configuration $cs\bar q \bar q$ and quantum numbers $I(J^P)=0(1^+)$ both in the single and coupled channels.
The current experimental information that we have in this sector comes from the $T_{cs/c\bar{s}}(2900)$ resonances. In order to produce some results, we first use the set of parameters used in~\cite{Molina:2022jcd} to reproduce these states within the hidden gauge approach in agreement with both available lattice data and experimental data. Thus, we fix $\gamma=1$~\cite{Gil-Dominguez:2023huq,Molina:2022jcd} and $\Lambda=1.1$~GeV~\cite{Molina:2022jcd}, while $\gamma^{\prime}$ is a free parameter.\footnote{{The factors $\gamma$ and $\gamma^{\prime}$ in Eq.~\eqref{Eq:potential_OME} quantify the flavor SU(4) breaking effect between the charmed and light mesons. The values of factors are one for the interaction involving only light mesons.}} We consider two cases to estimate the value of $\gamma^{\prime}$:
\begin{itemize}
    \item[(a)] {In~\cite{Molina:2010tx,Molina:2020hde}, the interaction for the $D^*\bar{K}^*$ system was studied and the $T_{cs}$ was dynamically generated from it. The $D^*\bar{K}^*$ state can decay into $D\bar{K}$ through the pion exchange in that framework. This decay was considered through a Feynman-box diagram where the $D^*D\pi$ vertex $g_{D^*D\pi}^\mathrm{exp}$ is along with a form factor $ e^{q^2_\pi/\Lambda^2}$, {where $q_\pi$ here refers to the four-momentum of the exchanged pion}, $g_{D^*D\pi}^\mathrm{exp}=8.95$ and $\Lambda=1,1.2~$GeV.  A similar interaction was used to generate the $T_{c\bar{s}}$ from the $D^*K^*$--$D_s^*\rho$ coupled-channel system in~\cite{Molina:2010tx,Molina:2022jcd}. In the present work, form factors are not considered, but instead we have an extra $\gamma'$ factor for the interaction. We can get the same results as in~\cite{Molina:2020hde,Molina:2022jcd} by taking $\gamma'=1.065$.}
    
    \item[(b)] {$\gamma'=2.82$. Once we match the decay width of $D^*\to D\pi$ from the heavy quark spin symmetry and flavor SU(4) symmetry, the flavor SU(4) breaking factor is determined to be $\gamma^{\prime}$=2.82, as detailed in Appendix~\ref{Sec:Dstar_decay}. }
\end{itemize}
We set SU(4) breaking factor $\gamma$ to 1, and consider the two sets of parameters (a) and (b). The possible molecules and the coupling constants between poles and relative channels can be extracted from the $T$-matrix in Eq.~\eqref{Eq:T_matrix}.

For the single channel case, as listed in Tables~\ref{Tab:pole_case1} and \ref{Tab:pole_case2}, and with both sets of parameters, (a) and (b), we observe two poles: a pole below the $D^*\bar K$ threshold, corresponding to the virtual state; the other pole around $D\bar K^*$ threshold represents {a bound state}. In the first case, the binding energy for the $D^*\bar K$ virtual state is {21.8 and 29.8} MeV for the sets (a) and (b), respectively. In the second case, for the $D\bar K^*$ bound state, {the binding energy is 0.6 MeV for both sets (a) and (b) due to the absence of $\gamma^{\prime}$ in the potential for the $D\bar K^*$ channel}.
The values of coupling $g_i$ in Tables \ref{Tab:pole_case1} and \ref{Tab:pole_case2} indicate that the $D$-wave contribution to the molecules is significantly smaller than that of $S$-wave. Therefore, these two states can be interpreted as $S$-wave states related to the $D^*\bar K$ and $D\bar K^*$ channels. Note that, for set (b), a bound state appears near the $D\bar K^*$ threshold due to the fact that the attractive interaction of the $\rho$ meson exchange in the $D\bar K^*$ channel is much stronger than that in $D^*\bar K$ channel. 

In the $D^*\bar K$-$D\bar K^*$ coupled-channel system, we find two poles in the unphysical RS, one close to $D^*\bar K$ and a higher one near the $D\bar K^*$ threshold. The lower pole is located below $D^*\bar K$ threshold about $17.7$ and $5.7$ MeV for the sets (a) and (b), respectively, as shown in Tables \ref{Tab:pole_case1} and \ref{Tab:pole_case2}. In set (a), this pole is strongly coupled to the $S$-wave $D^*\bar K$ channel, making the $D$-wave contributions for both channels minimal. While for the set (b), the large coupling of the coupled channel in Eq.~\eqref{Eq:potential_OME} significantly enlarges the coupling between this pole and $D\bar K^*$ threshold. The upper pole is a resonance with a splitting between the $D\bar K^*$ threshold and the real part of the pole position of 1.6 and {5.3} MeV for the sets (a) and (b), respectively. The coupling of this pole to the $S$-wave $D\bar K^*$ channel is larger than that of the other channels. This state can naturally decay into $D^*\bar{K}$. As shown in Table~\ref{Tab:pole_case2}, the coupling to the $D^*\bar K$ channel in $D$-wave is similar to that in the $S$-wave, and much smaller than that of $S$-wave $D\bar K^*$. The upper pole can decay to $D^*\bar K$ through both the scalar potential ($D^*_s$ exchange) and tensor interaction ($\pi$, $\eta$ and ~$\eta^{\prime}$ exchange). The scalar potential is suppressed by the mass of $D^*_s$, making the tensor interaction mostly responsible for the decay width of this state. {In the set (b), the decay width for the upper pole is greater than that of the set (a) due to the stronger coupled-channel effect.}
Comparing the coupled-channel case with the single-channel case, we find that the coupled-channel effect has minimal influence on the mass of the molecule (about {a few} MeV difference), but it has a significant impact on the width of the upper pole for sets (a) {and (b)}. 

\begin{table*}[!ht]
    \centering
    \renewcommand\arraystretch{1.6}
    \caption{Pole position and effective coupling of $D^*\bar K-D\bar K^*$ system. {We consider the SU(4) breaking factors as in set (a) $\gamma=1.0$, $\gamma^{\prime}=1.065$, and the cutoff $\Lambda=1.1$ GeV.} The poles for the single $D^*\bar K$ and $D\bar K^*$ system are listed in the second and third columns, respectively. The fourth and fifth columns display the poles for the $D^*\bar K-D\bar K^*$ coupled-channel system. The third row indicates the RS relevant to the two-body cut. All the poles are located at unphysical RS associated with $D\bar K \pi$ three-body cut.}\label{Tab:pole_case1}
\begin{ruledtabular}
    \begin{tabular}{l|c|c|c|c}
        & \multicolumn{2}{c|}{single channel} & \multicolumn{2}{c}{coupled channel}\\\hline
        Pole [MeV] & $ 2482.4$ & $2760.2-7.3i$ & $2486.5$ & $ 2759.2-8.9i$\\
        \hline
        RS & $(-)$ & $(+)$ & $(-,+)$ & $(-,+)$ \\ \hline\hline
        Channel&\multicolumn{4}{c}{Coupling $g_{i}$ [GeV] }
        \\\hline
        $D^*\bar K(S)$ & $4.10i$  &$ -$ &$3.96i$ &$ 0.23+0.40i$ \\ \hline
        $D^*\bar K(D)$ & $-0.00i$  &$-$ &$-0.00i$ &$ 0.32+0.04i $ \\ \hline
        $D\bar K^*(S)$ & $-$  &$ 1.89-0.52i $ &$2.25i$ &$  2.49+0.54i$\\ \hline
        $D\bar K^*(D)$ & $-$  &$  -0.00-0.00i$ &$ 0.17i$ &$ -0.01-0.02i$\\
    \end{tabular}
    \end{ruledtabular}
\end{table*}

 Our results show two isoscalar poles in the $D^*\bar K$-$D\bar K^*$ coupled-channel system, both around the relevant two-body thresholds. In comparison, two poles are also predicted with the on-shell approximation of the Bethe-Salpeter equation in Ref.~\cite{Gamermann:2007fi}, where the mass of the upper pole matches our findings.
However, due to the incorporation of momentum-dependent decay width for $\bar K^*$ in our analysis, the decay width of the upper pole is significantly smaller than the width predicted in Ref.~\cite{Gamermann:2007fi} once the $\bar K^*$ width is involved in the previous calculation. For the lower pole, the results in Ref.~\cite{Gamermann:2007fi} predict a broad cusp at the $D^*\bar K$ threshold, whereas our results suggest a virtual state below the $D^*\bar K$ threshold. This discrepancy in the decay width with respect to~\cite{Gamermann:2007fi} highlights the impact of including momentum-dependent decay widths in our calculations.

\begin{table*}[!ht]
    \centering
    \renewcommand\arraystretch{1.6}
    \caption{{Pole position and effective coupling of the $D^*\bar K-D\bar K^*$ system with the set (b) $\gamma=1$, $\gamma^{\prime}=2.82$, and $\Lambda=1.1$ GeV.} The RS for the poles is the same as that in Table \ref{Tab:pole_case1}. }\label{Tab:pole_case2}
    \begin{ruledtabular}
    \begin{tabular}{l|c|c|c|c}
        & \multicolumn{2}{c|}{single channel} & \multicolumn{2}{c}{coupled channel}\\\hline
        Pole [MeV] & $ 2474.4$ & $2760.2-7.3i$ & $2498.5$ & $ {2755.6-38.9i}$\\
        \hline
        RS & $(-)$ & $(+)$ & $(-,+)$ & ${(-,+)}$ \\ \hline\hline
        Channel&\multicolumn{4}{c}{Coupling $g_{i}$ [GeV] }
        \\\hline
        $D^*\bar K(S)$ & $4.21i$  &$ -$ &$3.13i$ & {$0.85+1.98i$} \\ \hline
        $D^*\bar K(D)$ & $-0.03i$  &$-$ &$-0.01i$ & {$ 1.30+0.00i $} \\ \hline
        $D\bar K^*(S)$ & $-$  &$ 1.89-0.52i $ &$4.60i$ & {$4.80+3.85i$}\\ \hline
        $D\bar K^*(D)$ & $-$  &$  -0.00-0.00i$ &$ 0.10i$ & {$ -0.11-0.21i$}\\
    \end{tabular}
    \end{ruledtabular}
\end{table*}

Furthermore, the behaviors of the two molecules can be observed in the $D^*\bar K$ invariant mass distribution. To guide the search for those two molecular states, we predict the line shape for the $D^*\bar K$-$D\bar K^*$ coupled-channel system, as shown in Fig. \ref{Fig:line_shape}.
Assuming that the production rate for the $D^*\bar K$ channel is much larger than that of $D\bar K^*$ channel, the amplitude $T_{11}(E)$ dominates the production amplitude. Then, the lower pole manifests as a cusp near the $D^*\bar{K}$ threshold for both parameter sets. {The line shapes associated with the upper pole exhibit a dip for both the sets (a) and (b)}. This is caused by the strong interaction in the cross channel $D^*\bar{K} \to D\bar{K}^*$, as discussed in Ref.~\cite{Dong:2020hxe}. Otherwise, when the production rate for $D\bar K^*$ is much higher than that of $D^*\bar K$ channel, a peak is present around $D\bar K^*$ threshold for the {sets (a) and (b)}, {as shown in Fig.~\ref{Fig:line_shape}}. In the future, the observation of the line shape in the $D^*\bar K$ invariant mass distribution will help us to search for those two molecular states and explore the production rates of those two molecules. 

\begin{figure*}
\centering	\includegraphics[scale=0.5]{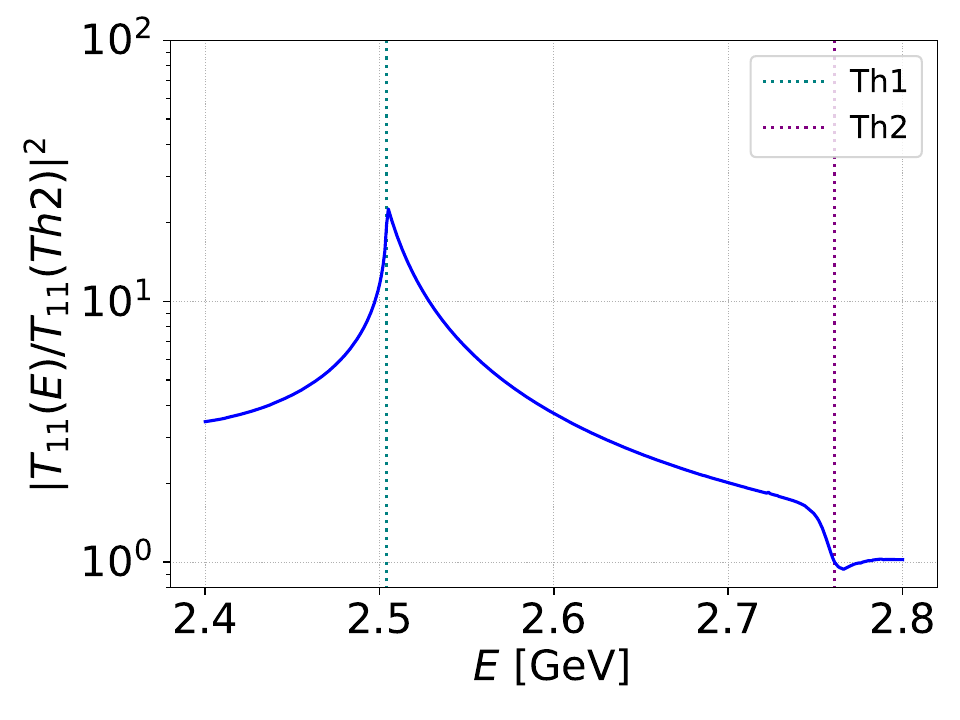}
 \includegraphics[scale=0.5]{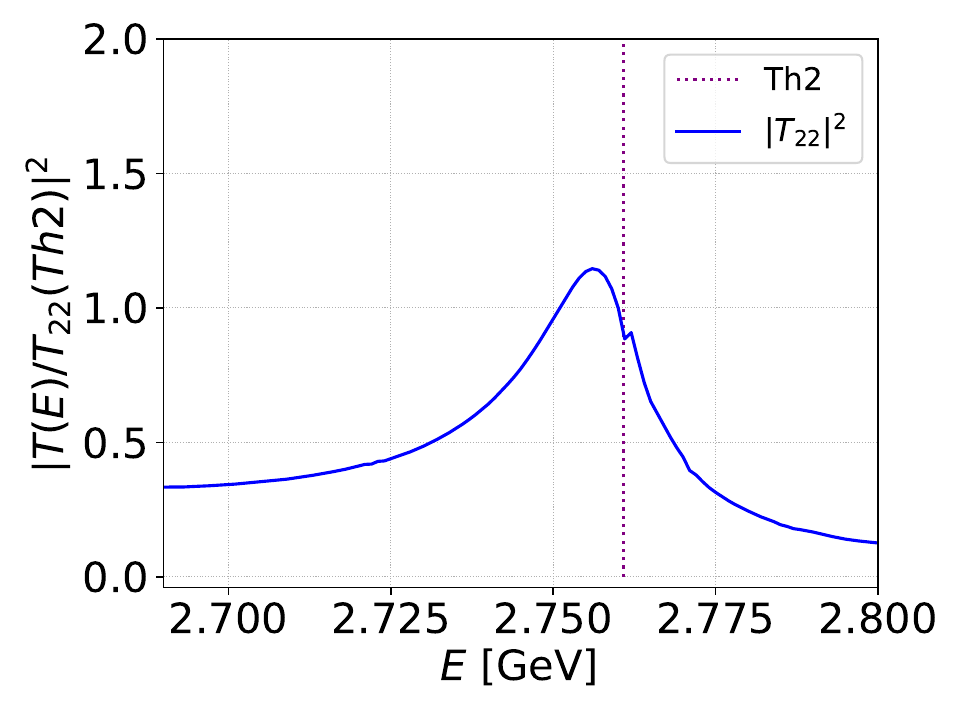}
\caption{{Line shapes for the $D^*\bar K-D\bar K^*$ coupled channel. Here Th1 and Th2 denote the $D^*\bar K$ and $D\bar K^*$ thresholds. Left: Line shape associated to the cusp around the $D^*\bar{K}$ threshold. Right: Line shape associated to the resonance found around the $D\bar{K}^*$ threshold. Both the left and right panels are associated with the set of parameters (a).} }
\label{Fig:line_shape}
\end{figure*}

\section{Summary}\label{sec:sum}

In the open-charm sector, the exotic structure $T_{cs}(2900)$ close to the $D^*\bar{K}^*$ {threshold} has been observed by the LHCb Collaboration~\cite{LHCb:2020bls,LHCb:2020pxc}. Given its quark content $cs\bar q \bar q$, this is a flavor exotic state. Especially, for the $D^*\bar K$-$D\bar K^*$ coupled-channel system, we expect similar structures close to the {$D^*\bar K$ and $D\bar K^*$ thresholds}, where the presence of the $D\bar K \pi$ three-body cut can have a role for the {state near $D\bar K^*$ threshold,} 
being responsible for its decay width. These thoughts have inspired this work.

In this work, we have derived the effective potential of $D^*\bar K$-$D\bar K^*$ coupled-channel system through the effective Lagrangian. Our approach incorporates pseudoscalar and vector mesons exchange, and encompasses three-body effects coming from the finite decay width of unstable meson $\bar K^*$ and the possibility of the exchanged pion being on-shell for the resonance close to the $D\bar{K}^*$ threshold decaying into $D\bar{K}\pi$. By solving the integral LS equation, we predicted the possible molecules in this system.

Based on our calculation, we predict two hadronic molecules in the $D^*\bar K$-$D\bar K^*$ coupled-channel system. One pole, located below the $D^*\bar K$ threshold, represents a $D^*\bar K$ virtual state, while the other pole corresponds to the $D\bar K^*$ resonance. 
The results obtained here should motivate the search for these states in the experiment, which, after the discovery of the $T_{cs/c\bar{s}}(2900)$ states.

\begin{acknowledgements}

P.-P.S. is grateful to Feng-Kun Guo, José Antonio Oller, and Mao-Jun Yan for useful discussions. R.~Molina acknowledges support from the ESGENT program with Ref. ESGENT/018/2024 and the PROMETEU program with Ref. CIPROM/2023/59, of the Generalitat Valenciana, and also from the Spanish Ministerio de Economia y Competitividad and European Union (NextGenerationEU/PRTR) by the grant with Ref. CNS2022-136146 and Ref. PID2023-147458NB-C21. This project has received funding from the European Union Horizon 2020 research and innovation program under the program H2020-INFRAIA-2018-1, grant agreement No. 824093 of the STRONG-2020 project.

\end{acknowledgements}

\appendix

\section{Partial wave projection of the meson-exchange potential}\label{Sec:partial_wave}

To calculate the partial-wave potential, we consider a two-body state with the quantum numbers labeled by $|LS,JM\rangle$, where $J~(M)$ is total angular momentum (its third component), $L$ is the orbital momentum and $S$ is the total spin. In two-body system, the orbital momentum $l$ and its third component $m$ is defined as
\begin{align}
\left|l m, s_1 s_2\right\rangle\equiv\frac{1}{\sqrt{4 \pi}} \int d \hat{p} Y_{l}^m(\hat{p})\left|\vec{p}, s_1 s_2\right\rangle,
\end{align}
where $s_1$ and $s_2$ are the spins of two particles, $\vec{p}$ is the center-of-mass (c.o.m) three-momentum, and $\hat{p}=\vec{p}/|\vec{p}|$.
The two-body state is written as
\begin{align}
 \left|LS, J M\right\rangle=\frac{1}{\sqrt{4 \pi}} \sum_{\substack{\sigma_1, \sigma_2 \\
\mu, m }} \int d \hat{p} Y_{L}^m(\hat{p})\left(\sigma_1 \sigma_2 \mu \mid s_1 s_2 S\right)(m \mu M \mid L S J) \left|\vec{p}, \sigma_1 \sigma_2\right\rangle,
\end{align}
where Clebsch-Gordan coefficient $(m_1m_2m_3|j_1j_2j_3)$ is relevant to the structure $j_1+j_2=j_3$ and $m_i$ the third components of $j_i$.
For the two-body scattering, the partial-wave amplitude can be expressed as~\cite{Gulmez:2016scm,Oller:2019rej}
\begin{align}
T_{L S ; \bar{L} \bar{S}}^{J}=&\frac{Y_{\bar{L}}^0(\hat{z})}{2 J+1} \sum_{\substack{\sigma_1, \sigma_2, \bar{\sigma}_1 \\
\bar{\sigma}_2, m}} \int d \hat{p}^{\prime \prime} Y_{L}^m\left(\hat{p}^{\prime \prime}\right)^*\left(\sigma_1 \sigma_2 M \mid s_1 s_2 S\right)(m M \bar{M} \mid L S J)\left(\bar{\sigma}_1 \bar{\sigma}_2 \bar{M} \mid \bar{s}_1 \bar{s}_2 \bar{S}\right)\nonumber\\
&\times(0 \bar{M} \bar{M} \mid \bar{L} \bar{S} J)
\left\langle\vec{p}^{\,\prime \prime}, \sigma_1 \sigma_2|\hat{T}||\vec{p}| \hat{z}, \bar{\sigma}_1 \bar{\sigma}_2\right\rangle,
\end{align}
where $\vec{p},~\bar{L}$ and $\bar{S}$ ($\vec{p}\,^{\prime \prime},~L$ and $S$) denote the c.o.m. three-momentum, orbital momentum and spin for the initial (final) states, respectively. Notice that the c.o.m three-momentum of the initial states is along the $z$ axis.

With the use of such technique, we can evaluate the partial-wave potentials. For the potentials in Eq.~\eqref{Eq:potential_OME}, the Mandelstam variables $t$ and $u$ are the function of $\text{cos}~\theta$ (labeled by $z$ in the following), where $\theta$ is the angle between the momenta $p_1$ and $k_1$. To simplify the expression, we just keep the part of potentials associated to the polarization vectors, and the rest part is denoted by $D(s,u,t)$\footnote{Without on-shell approach, the momenta is off-shell and thus $s+u+t\neq \sum_i m_i^2$. Consequently, the meson-exchange potential is the function of $s$, $t$, and $u$.}. For the potential $V=D(s,t,u)\varepsilon(p_1)\cdot\varepsilon^*(k_1)$, the contributions of the $S$- and $D$-waves are
\begin{align}
V^{SS}(p,k)&=\int_{-1}^{1}dz~D(s,t,u)\left[ -\frac{1}{2}+\frac{pkz}{6m_{p_1}m_{k_1}}\right],\nonumber\\
V^{SD}(p,k)&=-\sqrt{2}\int_{-1}^{1}dz~D(s,t,u)\left[\frac{pkz}{6m_{p_1}m_{k_1}}\right],\nonumber\\
V^{DS}(p,k)&=-\sqrt{2}\int_{-1}^{1}dz~D(s,t,u)\left[\frac{pkz}{6m_{p_1}m_{k_1}}\right],\nonumber\\
V^{DD}(p,k)&=\int_{-1}^{1}dz~D(s,t,u)\left[+
\frac{1}{4}+
\frac{pkz}{3m_{p_1}m_{k_1}}{ -}\frac{3z^2}{4}\right],
\label{Eq:epsilon_p1_k1}
\end{align}
where we take the approach $E_{i}/m_{i}\simeq 1$, and the three-momenta $p=|\vec p_1|$ and $k=|\vec k_1|$ in the c.o.m frame.
The potential with polarization $\varepsilon(p_2)\cdot\varepsilon^*(k_2)$ leads to the same form of partial-wave potential as in Eq. \eqref{Eq:epsilon_p1_k1}, but with the masses of the initial and final states exchanged, i.e. $m_{p_1(p_2)}\to m_{p_2(p_1)}$ and $m_{k_1(k_2)}\to m_{k_2(k_1)}$.
For the potential with the structure $\varepsilon(p_1)\cdot\varepsilon^*(k_2)$, the contributions of the $S$- and $D$-waves are
\begin{align}
V^{SS}(p,k)&=\int_{-1}^{1}dz~D(s,t,u)\left[{ -}\frac{1}{2}-\frac{pkz}{6m_{p_1}m_{k_2}}\right],\nonumber\\
V^{SD}(p,k)&=\sqrt{2}\int_{-1}^{1}dz~D(s,t,u)\left[\frac{pkz}{6m_{p_1}m_{k_2}}\right],\nonumber\\
V^{DS}(p,k)&=\sqrt{2}\int_{-1}^{1}dz~D(s,t,u)\left[\frac{pkz}{6m_{p_1}m_{k_2}}\right],\nonumber\\
V^{DD}(p,k)&=\int_{-1}^{1}dz~D(s,t,u)\left[{ +}
\frac{1}{4}-
\frac{pkz}{3m_{p_1}m_{k_2}}{ -}\frac{3z^2}{4}\right].
\label{Eq:epsilon_p1_k2}
\end{align}

The potential relevant to the tensor force can be denoted by $V=D(s,t,u)k_2\cdot\varepsilon(p_1)p_2\cdot\varepsilon^*(k_1)$. The partial-wave potentials are expressed as
\begin{align}
V^{SS}(p,k)&=\int_{-1}^{1}dz~D(s,t,u)\left[\frac{k^2\mpw}{6\mko}+\frac{p^2\mkw}{6\mpo}+\frac{pkz}{6}\left(1+\frac{\mpw\mkw}{\mpo\mko}\right)\right],\nonumber\\
V^{SD}(p,k)&=-\sqrt{2}\int_{-1}^{1}dz~D(s,t,u)\left[\frac{k^2\mpw}{6\mko}-\frac{p^2\mkw}{12\mpo}+\frac{pkz}{6}\left(1+\frac{\mpw\mkw}{\mpo\mko}\right)+\frac{p^2z^2\mkw}{4\mpo}\right],\nonumber\\
V^{DS}(p,k)&=-\sqrt{2}\int_{-1}^{1}dz~D(s,t,u)\left[\frac{p^2\mkw}{6\mpo}-\frac{k^2\mpw}{12\mko}+\frac{pkz}{6}\left(1+\frac{\mpw\mkw}{\mpo\mko}\right)+\frac{k^2z^2\mpw}{4\mko}\right],\nonumber\\
V^{DD}(p,k)&=\int_{-1}^{1}dz~D(s,t,u)\left[-\frac{k^2\mpw}{6\mko}-\frac{p^2\mkw}{6\mpo}-\frac{5pkz}{12}+\frac{pkz\mpw\mkw}{3\mpo\mko}+\frac{k^2z^2\mpw}{2\mko}\right.\nonumber\\
&\qquad\qquad\qquad\qquad\left.+\frac{p^2z^2\mkw}{2\mpo}+\frac{3pkz^3}{4}\right],
\label{Eq:k2_p1}
\end{align}
while the partial-wave potentials for the tensor structure $k_1\cdot\varepsilon(p_1)p_2\cdot\varepsilon^*(k_2)$ are
\begin{align}
V^{SS}(p,k)&=\int_{-1}^{1}dz~D(s,t,u)\left[\frac{p^2\mko}{6\mpo}+\frac{k^2\mpw}{6\mkw}-\frac{pkz}{6}\left(1+\frac{\mpw\mko}{\mpo\mkw}\right)\right],\nonumber\\
V^{SD}(p,k)&=-\sqrt{2}\int_{-1}^{1}dz~D(s,t,u)\left[\frac{k^2\mpw}{6\mkw}-\frac{p^2\mko}{12\mpo}-\frac{pkz}{6}\left(1+\frac{\mpw\mko}{\mpo\mkw}\right)+\frac{p^2z^2\mko}{4\mpo}\right],\nonumber\\
V^{DS}(p,k)&=-\sqrt{2}\int_{-1}^{1}dz~D(s,t,u)\left[\frac{p^2\mko}{6\mpo}-\frac{k^2\mpw}{12\mkw}-\frac{pkz}{6}\left(1+\frac{\mpw\mko}{\mpo\mkw}\right)+\frac{k^2z^2\mpw}{4\mkw}\right],\nonumber\\
V^{DD}(p,k)&=\int_{-1}^{1}dz~D(s,t,u)\left[-\frac{k^2\mpw}{6\mkw}-\frac{p^2\mko}{6\mpo}+\frac{5pkz}{12}-\frac{pkz\mpw\mko}{3\mpo\mkw}+\frac{k^2z^2\mpw}{2\mkw} \right.\nonumber\\
&\qquad\qquad\qquad\qquad\left.+\frac{p^2z^2\mko}{2\mpo}-\frac{3pkz^3}{4}\right].
\label{Eq:k1_epsilon}
\end{align}

\section{Decay for $D^*\to D\pi$}\label{Sec:Dstar_decay}
{

Since we have used SU(4) symmetric lagranigans, and break the symmetry with the masses and breaking factors $\gamma'$, it is useful to compare the result for the $\Gamma(D^*\to D\pi)$ decay width from these Lagrangians, with the one obtained from a Lagrangian based on Heavy Quark Symmetry (HQS). According to HQS, the leading order covariant chiral Lagrangian for $D^*D\pi$ is~\cite{Casalbuoni:1996pg,Casalbuoni:1992gi}
\begin{align}
{\cal L}_{HQ} =& \frac{\sqrt{2}ig_2}{F_{\pi}}\left(P_{a,\mu}^{*(Q)}{}^{\dag}P_b^{(Q)} + P_{a,\mu}^{*(Q)}P_b^{(Q)}{}^{\dag} -i\epsilon^{\mu\nu\alpha\beta}v_{\beta}P_{a,\alpha}^{*(Q)}{}^{\dag}P_{a,\beta}^{*(Q)} \right)\partial^{\mu}\phi_{ab},
\label{Eq:Lag_chpt_H}
\end{align}
with $P^{*(Q)} = \left\{D^{*0},D^{*+}\right\}$, $P^{(Q)} = \left\{D^{0},D^{+}\right\}$, and $v_{\alpha}$ the four-velocity of $P^*$. The pseudoscalar field $\phi$ is 
\begin{align}
\phi=\left(\begin{array}{cc}
\frac{\pi^{0}}{\sqrt 2} & \pi^{+} \\
\pi^{-} & -\frac{\pi^{0}}{\sqrt 2}
\end{array}\right).
\end{align}

In the frameworks of flavor SU(4) and heavy quark spin symmetry, the decay widths for $D^{*+}(p)\to D^0(k)\pi^+(q)$ calculated using the Lagrangians in Eqs. \eqref{Eq:Lagrangian_SU4} and \eqref{Eq:Lag_chpt_H}, are given by 
\begin{align}
\Gamma_{\text{SU(4)}}[D^{*+}\to D^0\pi^+]&= \frac{|\vec q|}{24\pi m_{D^*}^2}\left(\gamma^{\prime}g\right)^2\frac{\lambda(m_{D^*}^2,m_D^2,m_{\pi}^2)}{m_{D^*}^2},\\
\Gamma_{\text{HQ}}[D^{*+}\to D^0\pi^+]&= \frac{|\vec q|}{24\pi m_{D^*}^2}\frac{g_2^2}{F_{\pi}^2}\frac{\lambda(m_{D^*}^2,m_D^2,m_{\pi}^2)}{2m_{D^*}^2}m_{D^*}m_{D}\ ,
\end{align}
respectively. By matching these two formulas, the SU(4) breaking factor $\gamma^{\prime}$ obtained is
\begin{align}
\gamma^{\prime} = \frac{\sqrt {2m_{D^*}m_{D}}g_2}{m_V}.
\end{align}
We determine the effective coupling $g_2=0.566$ from the experimentally partial width of $D^{*+} \to D^0\pi^+$ \cite{ParticleDataGroup:2022pth}. Therefore, this factor is $\gamma^{\prime}=2.82$.
}

\bibliography{DstarKbar.bib}
\end{document}